# Simulating the phase behavior of the Kuramoto tree

Mohammad Javad Nouhi, Javad Noorbakhsh

## Abstract:

The Kuramoto model is a versatile mathematical framework that explains phenomena resulting from interactions among phase oscillators. It finds applications in various scientific and engineering domains. In this study, we focused on a Y-shaped network, which serves as the fundamental unit of a tree network. By simulating oscillators on the network, we generated heat maps for different numbers of nodes and coupling strengths and demonstrated the occurrence of different phases. Our findings reveal transitions between synchronization, wave state, and chaos within the system.

## Introduction:

The Kuramoto model is a mathematical model used to simulate oscillatory behavior in a complex system, and is based on a system of coupled differential equations that describe the behavior of a network of phase oscillators. This model has been applied to a wide range of problems in biology, physics, electrical engineering, and a myriad of other fields [1]; including but not limited to the regulation of circadian rhythms in the brain [2], oscillatory behavior in gene regulatory networks [3] [4], frequency synchronizations across a network of generators on a power grid [5], and synchronized flashing of firefly populations [6].

The dynamics and phases of Kuramoto oscillators on networks have been studied extensively [7, 8, 9], however, such studies mostly focus on fully and randomly connected networks. Some authors have explored other network structures such as trees. For example Dekkar, et al [10] provide a comprehensive study of Kuramoto oscillators on a tree in the synchronized regime, but overall the topic remains less explored due to a lack of closed-form equations. Nevertheless, these structures can arise in many real-life systems where three or more populations of oscillators interact.

In this paper we study a Y-shaped network, which is the simplest form of the Kuramoto tree. We show that despite the simplicity of this structure, it exhibits a rich range of behaviors, including synchronization, as well as chaotic, and wave-like behavior.

## Results :

Each Kuramoto oscillator is characterized by a phase variable $\theta_i$ which is coupled to other oscillators through the following equation:

$$\frac{d\theta_i}{dt} = \omega_i + \frac{K}{N}\sum_{j} Sin(\theta_i - \theta_j)$$

Here the index i corresponds to the i'th oscillator. The $\omega_i$ is the natural frequency of oscillator i, K is the coupling constant and N is the number of oscillators. [11].

Phase-order parameter r is defined as follows:

$$r = \left| \frac{1}{N}\sum_{j=1}^{N} e^{i\theta_j} \right|$$

In this work, we built a coupled system of Kuramoto oscillators on a Y-shaped network (Fig 1) with a main branch with 10*n nodes (n ranging from 0 to 9) and two branches of length 5.

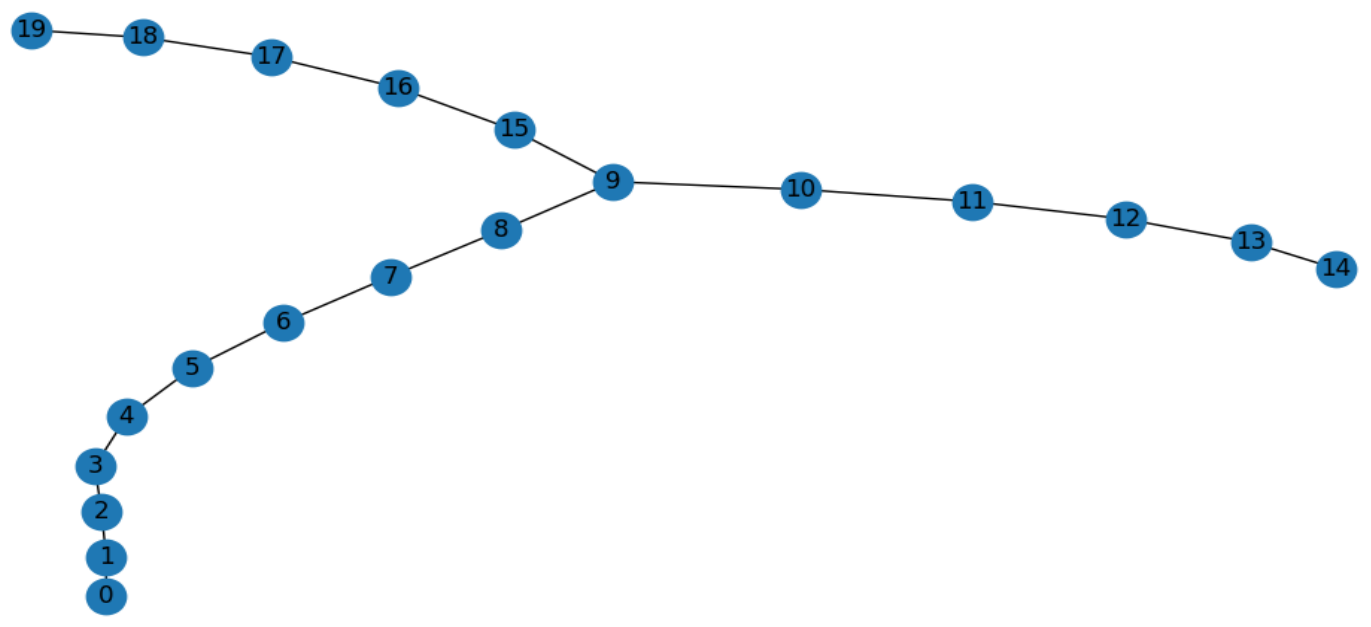

Figure 1. Example Y-shaped network used in the simulations with n=20 nodes. Each node is a Kuramoto oscillator and the links correspond to non-zero coupling constants.

We explored the phases of the model in several parameter regimes corresponding to different natural frequencies and coupling constants.
In particular, we simulated the system by choosing the frequency of the oscillators [ω] from three different distributions:

1. Constant frequency: delta Dirac distribution
2. Narrow spread: Normal random distribution with standard deviation of 0.1
3. Wide spread: Normal random distribution with standard deviation of 1.0

To systematically explore the behavior of this system we exploit two different order parameters. We define an order parameter which corresponds to the time to reach synchronization (see Methods) and is based on the time required for the phase-order parameter r to reach the threshold 0.9. This order parameter can aid us in pinpointing phase transitions to/from synchronization. We also take advantage of a Fourier-based order parameter to assess wave-like behavior (see Methods). In what follows, we will describe the results for each choice of frequency distribution.

First we simulated the model with the constant frequency and examined its qualitative behavior using the heatmap of the order parameter for different number of nodes and coupling strengths and inspected individual time-course plots for select points on the phase diagram (Figure 2). Overall, we observe a decrease in the minimum time to synchronization as the coupling strength increases for a fixed node.

Whereas for a fixed coupling strength with increasing the nodes the minimum time for synchronization increases up to a maximum around N=60 and then decreases as N is increased as well ( Figure 2).

a)

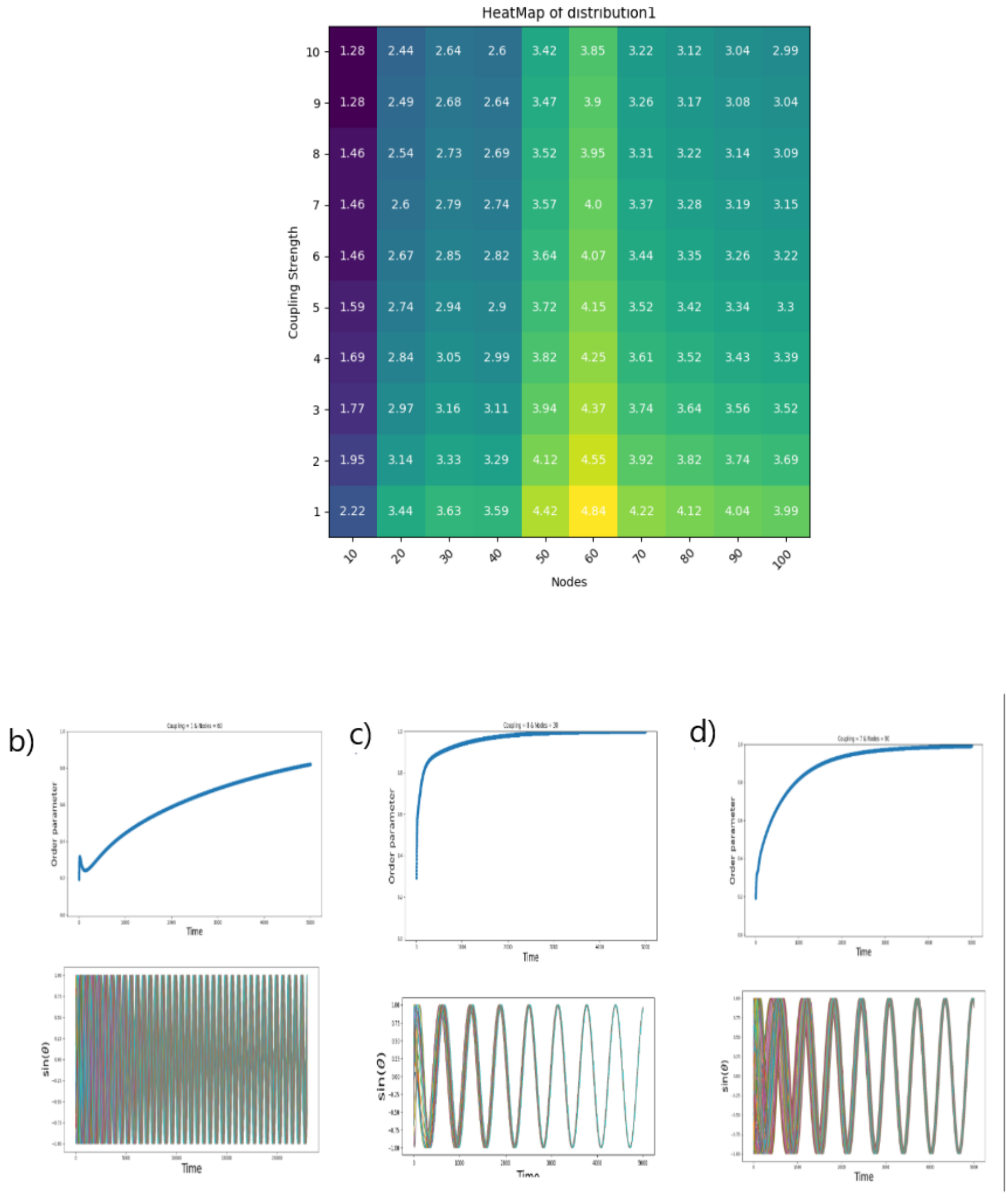

**Figure 2**. **Constant frequency phases. a)** Heat map of the time to synchronization .Values are the logarithm of the time taken to reach the order parameter to 0.9. Our simulation time is 5000 units. Initial conditions were set to normal random. Phase-Order parameter and population time course plot for **b)** for Nodes = 60 and Coupling strength = 1 **c)** Nodes = 30 and Coupling strength = 8 **d)** Nodes = 90 and Coupling strength = 7.

We next perturbed the system parameters by selecting the natural frequencies from a normal random distribution with a small standard deviation (σ=0.1). Similar to the previous case, we examined its qualitative behavior using time-course plots and heatmap of nodes and coupling strengths (Fig 3). Parameters for which the system did not reach synchronization are left blank in the heatmap exhibiting a richer behavior compared to the constant frequency simulations. The empty states are out of phase synchronized .
We observe a decrease in the minimum time to synchronization as the coupling strength increases for a fixed number of nodes. Whereas for a fixed coupling strength with increasing the number of nodes the minimum time for synchronization increases up to a maximum around N=70 and then decreases as N is increased as well (Fig 3).

a)

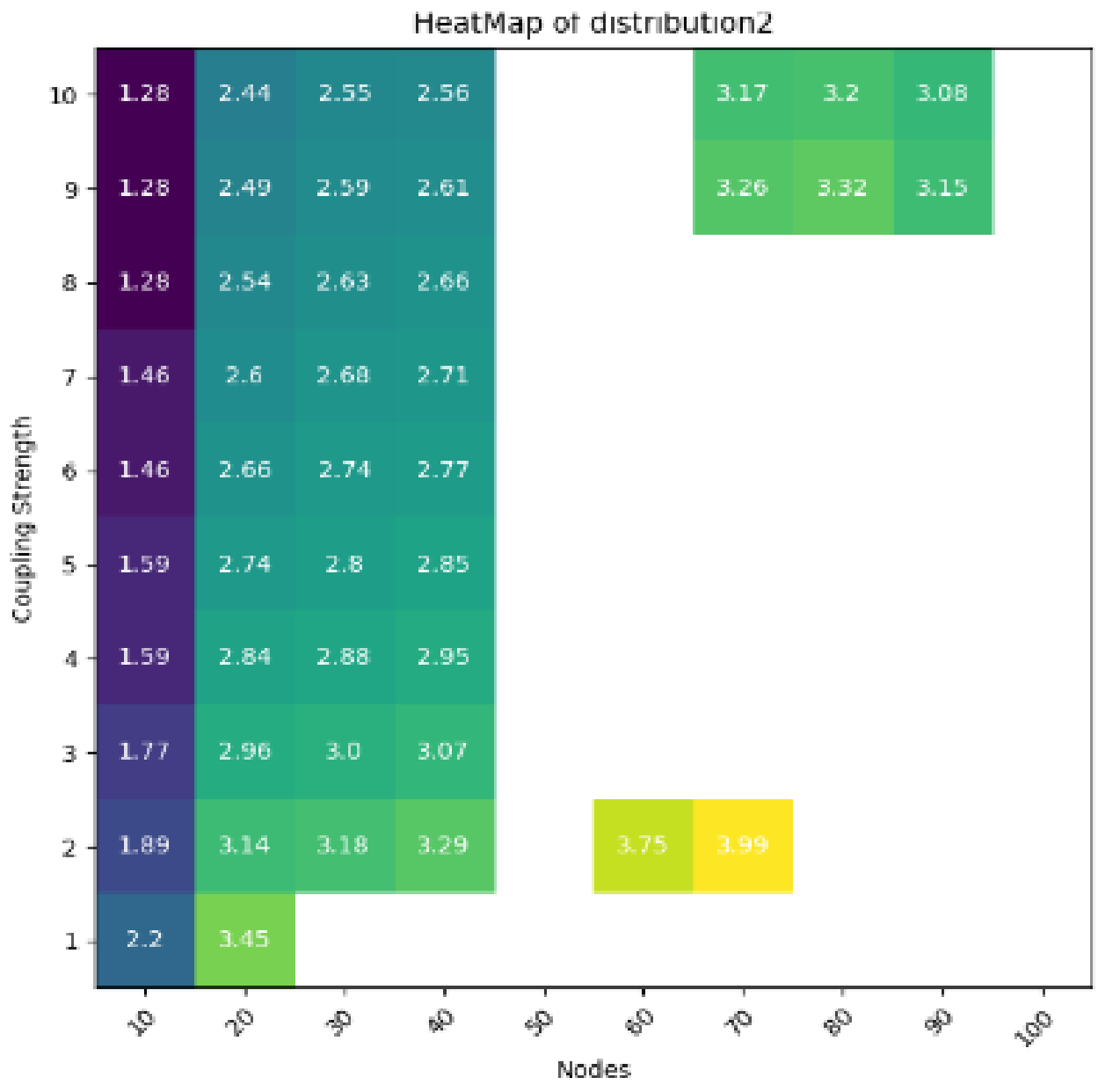

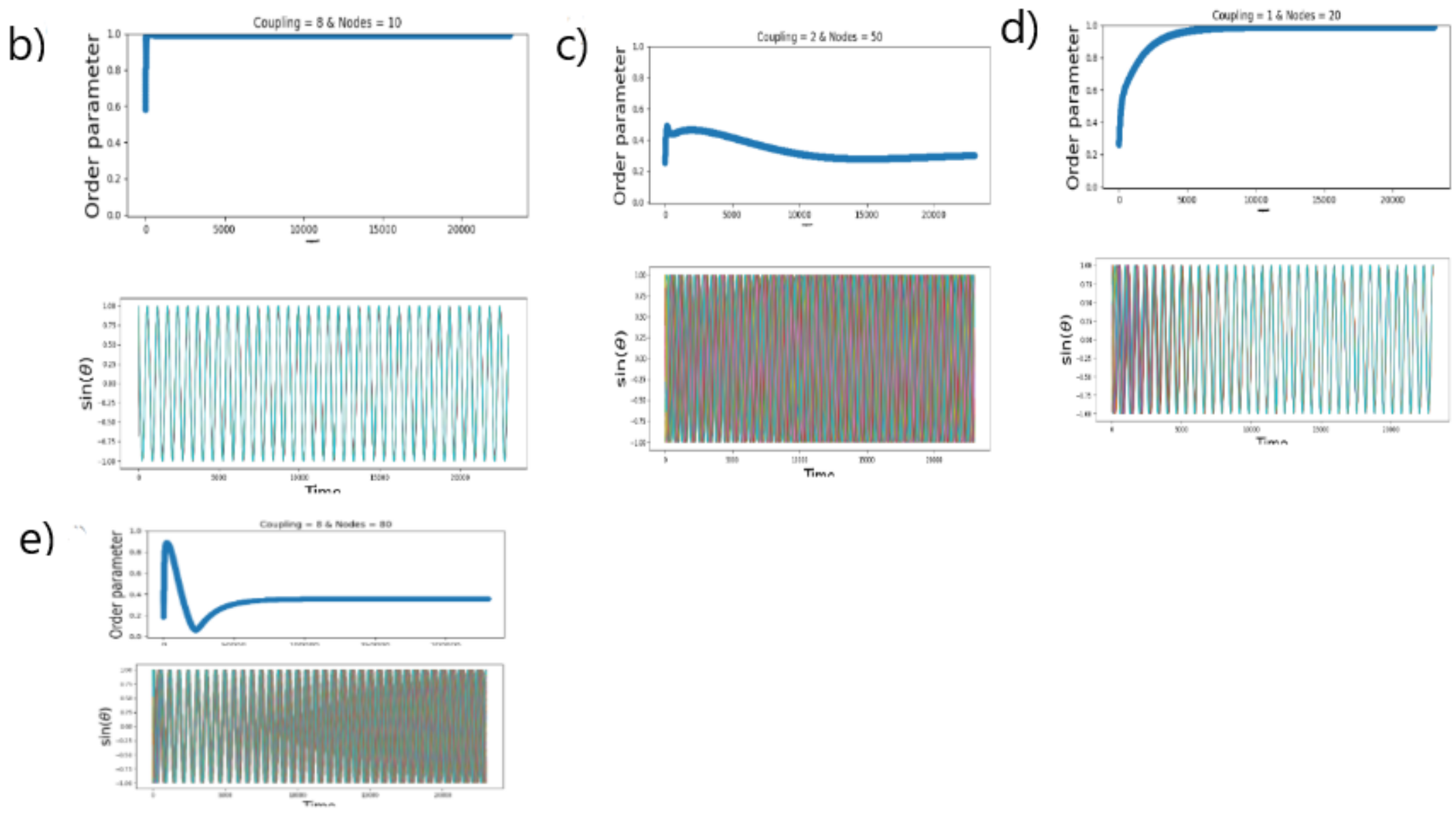

**Figure 3. Narrow spread. a)** Heat map of the time to synchronization .Values are the logarithm of the time taken to reach the order parameter to 0.9. Our simulation time is 5000 units. Initial conditions were set to normal random. Phase-Order parameter and population time course plot for **b)** for Nodes = 10 and Coupling strength = 8 **c)** Nodes = 50 and Coupling strength = 2 **d)** Nodes = 20 and Coupling strength = 1 **e)** Nodes = 80 and Coupling strength = 8**.**

Finally, we further perturbed the system parameters by selecting the natural frequencies from a normal random distribution with a larger standard deviation (σ=1.0). Upon careful inspection of the time course plots we noticed a more complex landscape of behaviors. Therefore, we modified our phase diagram to include three different qualitative phases, namely 1. Chaotic state, 2. Wave state, and
3. Synchronized state. We further determined these states by inspecting the behavior of the system in the frequency domain using the fast fourier transform (FFT) of phase-order parameter time courses (see Methods). . Phase diagram of the system corresponding to these three states is shown in Figure 4 alongside select time-courses and their corresponding frequency plots.

The system undergoes several transitions, in particular increasing the number of nodes and decreasing the coupling strength leads to chaotic behavior. Interestingly the synchronized behavior is unique to the region with low numbers of nodes and high coupling constant, while the oscillatory regime tends to appear as intermediary between the other two phases (Figure 4).

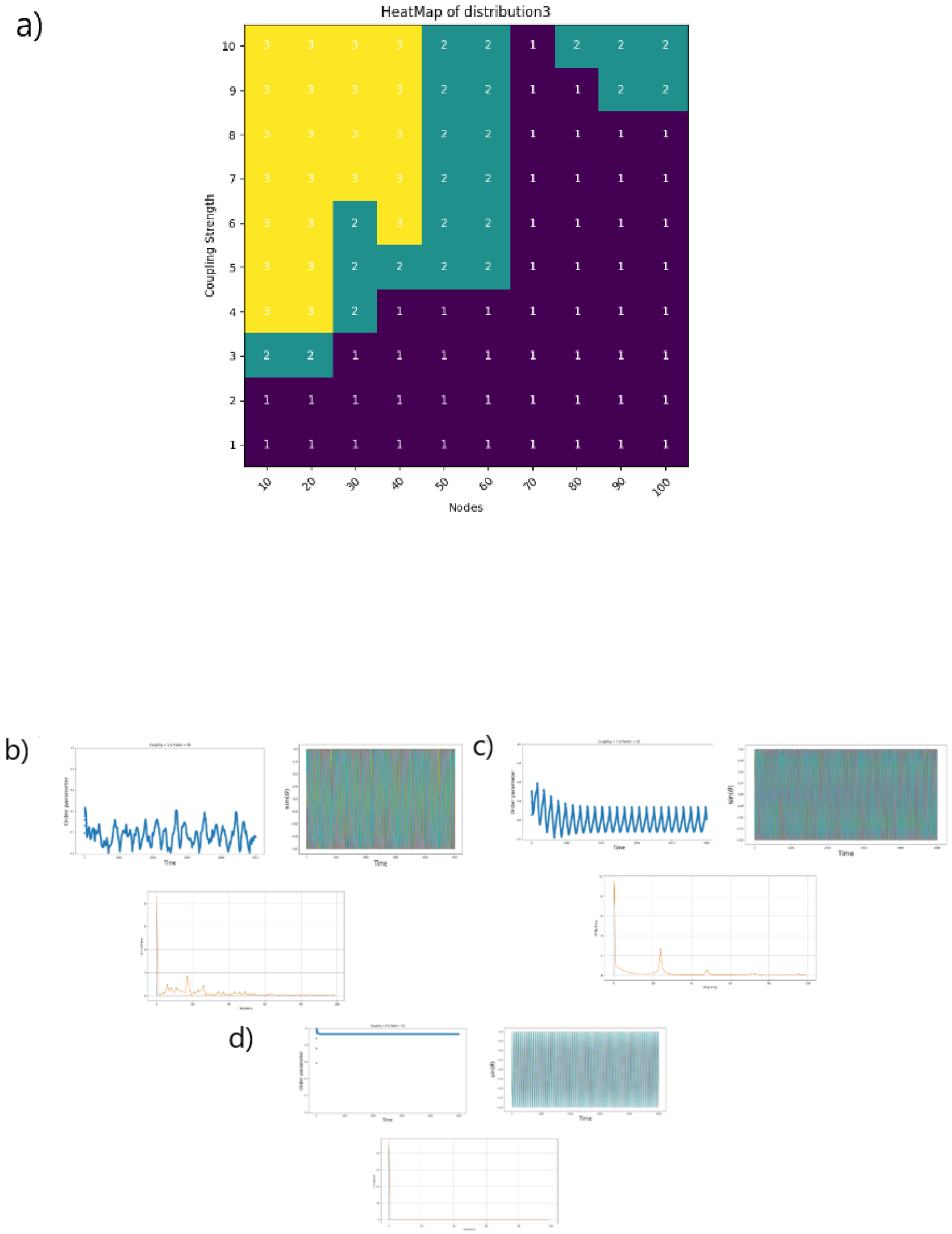

**Figure 4. Wide spread. a)** Heat map of the time to synchronization .1 or 2 or 3 Values are the states of the order parameter time course plot. Our simulation time is 5000 units. Initial conditions were set to normal random. Phase-Order parameter and population time course plot and FFT plot **b)** for "1" or Chaotic state **c)** for "2" or Wave state **d)** for "3" or Synchronized state.

## Discussion:

Despite its simplicity, the Kuramoto model plays an important role in the characterization and modeling of the dynamics in many different systems with coupled oscillators. The combinations of different topologies and variations of the model dynamics are countless, satisfying the needs in scientific domains and problem settings.

In this paper, we studied the Kuramoto oscillators on a Y-shaped network and explored their properties in a few parameter regimes.We showed that complex and potentially chaotic behavior can arise as the system parameters are perturbed away from a uniform model of identical oscillators. The degree of this randomness can be controlled by the width of the normal distribution, $\sigma$, from which natural frequencies of the oscillators are drawn. Increasing this parameter pushes the system out of a globally synchronized state into wave-like and chaotic phases.

In this work we only focused on a Y-shaped network with Gaussian parameter distribution. Extensions of this model can be considered where more complex tree-like structures with multiple leaves are used and parameters are drawn from different distributions with spatial correlations. Furthermore, we only focused on a single snapshot of the system parameters and a single random initial condition. An ensemble of such parametrizations would be needed to fully understand the tendency of the system to fall into different qualitative regimes.

Overall, we show that the Kuramoto model on a simple network can exhibit a rich range of behaviors which may have manifestations in natural coupled oscillators.

## Methods:

**Simulations:** Simulations were done using the Python library Kuramoto modified to mimic the network discussed in this manuscript.

**Time to synchronization order parameter:** We defined the time-point at which r (the phase-coherence of oscillators) reaches 0.9 as a threshold for synchronization of the system. Heat maps in figures 2 and 3 were plotted using this quantity, where its logarithm was shown to shrink the dynamic range of the values.

**Phase diagram using frequency domain:** The phase diagram in figure 4, was produced by taking the Fast Fourier Transform (FFT) of the time course of r and identifying its peaks.

The system's phase was then determined based on the number and structure of the peaks. Each point on the diagram was assigned to one of three phases of chaotic, wave, and synchronized. The Fourier transform was applied to the order parameter from t=0 to t=5000, using the FFT function from the scipy package in Python.

The peaks were determined by using the peak_finds function from the scipy library. We defined the state of the system according to the number of the peaks, with no peaks corresponding to synchronization.

For cases with at least 1 peak but less than 5 total peaks we assigned the wave attribute. Careful visual inspection revealed extra harmonics in this case. The cases with more than 5 peaks were assigned to the chaotic category.